\documentclass[twocolumn]{aastex631}

\usepackage{amsmath}

\newcommand\pp{PSR~B0540$-$69}

\shorttitle{The irregular rotation of PSR B0540$-$69}
\shortauthors{Espinoza et al.}
\graphicspath{{./}{figures/}}

\begin{document}

\title{A growing braking index and spin-down swings for the pulsar PSR B0540$-$69}

\correspondingauthor{Crist\'obal M. Espinoza}
\email{cristobal.espinoza.r@usach.cl}

\author[0000-0003-2481-2348]{Crist\'obal M. Espinoza}
\affiliation{Departamento de F\'isica, Universidad de Santiago de Chile {\sc (USACH)},  Chile}
\affiliation{Center for Interdisciplinary Research in Astrophysics and Space Sciences {\sc (CIRAS)}, {\sc USACH}, Chile}

\author[0000-0002-7889-6586]{Lucien Kuiper}
\affiliation{SRON-Netherlands Institute for Space Research, The Netherlands}

\author[0000-0002-6089-6836]{Wynn C. G. Ho}
\affiliation{Department of Physics and Astronomy, Haverford College, USA}

\author{Danai Antonopoulou}
\affiliation{Jodrell Bank Centre for Astrophysics, Department of Physics and Astronomy, The University of Manchester, UK}

\author[0009-0008-6187-8753]{Zaven Arzoumanian}
\affiliation{X-Ray Astrophysics Laboratory, NASA Goddard Space Flight Center, USA} 

\author[0000-0001-6119-859X]{Alice K. Harding}
\affiliation{Theoretical Division, Los Alamos National Laboratory, USA} 

\author{Paul S. Ray}
\affiliation{Space Science Division, U.S. Naval Research Laboratory, Washington, USA}

\author[0000-0002-7991-028X]{George Younes}
\affiliation{Astrophysics Science Division, NASA Goddard Space Flight Center, USA}


\begin{abstract}

The way pulsars spin down is not understood in detail, but a number of possible physical mechanisms produce a spin-down rate that scales as a power of the rotation rate ($\dot\nu\propto-\nu^n$), with the power-law index $n$ called the braking index. 
\pp\ is a pulsar that in 2011, after 16 years of spinning down with a constant braking index of $2.1$, experienced a giant spin-down change and a reduction of its braking index to nearly zero. 
Here, we show that following this episode the braking index monotonically increased during a period of at least four years and stabilised at $\sim1.1$. 
We also present an alternative interpretation of a more modest rotational irregularity that occurred in 2023, which was modelled as an anomalous negative step of the rotation rate. 
Our analysis shows that the 2023 observations can be equally well described as a transient swing of the spin-down rate (lasting $\sim65$ days), and the Bayesian evidence indicates that this model is strongly preferred. 

\end{abstract}


\section{Introduction} \label{sec:intro}

The rotation rates of isolated pulsars decrease gradually over timescales that range from thousands to millions of years, due to electromagnetic radiation, plasma flows, and the emission of gravitational waves. 
The observed rates of loss of rotational kinetic energy, or spin-down luminosity $\dot{E}=-4\pi^2 I\nu\dot\nu$, cover more than eight orders of magnitude and can be as high as $10^{38}$\,erg\,s$^{-1}$ for the youngest pulsars. 
Above, $\nu$ is the spin frequency, $\dot\nu$ is its first time derivative ($\dot\nu<0$), and the moment of inertia is taken to have a typical value of $I=10^{45}$\,g\,cm$^{-2}$ \citep[e.g.,][]{2004hpa..book.....L}. It is customary to assume a power-law scaling for the spin-down $\dot{\nu}$ as
\begin{equation}
\label{powerlawspindown}
    \dot{\nu}=-k\nu^n  \quad ,
\end{equation}
in which $n$ is called the braking index and $k$ depends on stellar parameters. 
The braking index is expected to be $n\simeq3$ if electromagnetic radiation from a rotating dipole in vacuum dominates the spin-down, acceleration of particle winds will give $n\simeq1$, and emission of gravitational waves predicts values $n>3$ \citep{1999ApJ...525L.125H,2018ASSL..457..673G}.

The braking index $n$ can be estimated from observations as $n=\nu\ddot{\nu}/\dot{\nu}^2$, assuming $k$ constant, and provided the very small $\ddot{\nu}$ is reliably measured. 
Irregularities of the generally stable pulsar spindown occur in all pulsars to some degree and contaminate measurements of the long-term $\ddot{\nu}$. 
Timing noise, a continuous wandering of the rotation with respect to a simple spin-down model, is a widespread phenomenon and can cause the observed $\ddot{\nu}$ to vary widely from negative to positive values \citep{2010MNRAS.402.1027H}. 
In a number of cases, timing noise was found to consist of transitions between distinct states, characterised by different emission properties and a different spin-down rate $\dot\nu$ \citep{2010Sci...329..408L}. 
Other irregularities, more common to young pulsars, are spin-up glitches: sporadic, sudden, and very fast increases of the rotation rate ($\Delta\nu > 0$ changes) usually accompanied by an abrupt spindown rate decrease ($\Delta\dot{\nu}<0$) \citep{2022RPPh...85l6901A}. 
Glitches are followed by relaxation of the rotational parameters over timescales of weeks to years, and this recovery often induces high values of $\ddot{\nu}$ \citep{2024MNRAS.532..859L}. 
The few $\ddot{\nu}$ measurements available that might reflect the underlying, long-term behaviour indicate braking indices of different values and less than three \citep[e.g.,][]{2007Ap&SS.308..317L,2017MNRAS.466..147E,2020MNRAS.494.2012P}. 
A number of processes have been suggested to explain the observed $n<3$, from strong spindown contributions by energetic stellar winds (for which $n=1$), to magnetic field evolution or progressive decoupling of internal superfluid components \citep[e.g.,][]{1988MNRAS.234P..57B,2012NatPh...8..787H,2015MNRAS.452..845H,2017MNRAS.469.1974E,2017MNRAS.466..147E,2020MNRAS.494.1865W}.

PSR B0540$-$69 (PSR J0540$-$6919) is an energetic, young, X-ray pulsar in the Large Magellanic Cloud that rotates with a period of $50$\,ms ($\nu\sim20$ Hz). 
It exhibits the third largest known spin-down luminosity, $\dot{E}=2\times 10^{38}$\,erg\,s$^{-1}$, which is likely to be the energy source of a bright synchrotron nebula that surrounds the pulsar and is visible from the radio waves to X-rays \citep{2019NatAs...3.1122G,2024ApJ...962...92X}. 
Considering only electromagnetic dipole radiation (for which $n=3$), the orthogonal component of the dipolar magnetic field at the surface is found as $B=3.2\times10^{19}\sqrt{-\dot\nu/\nu^3}$. 
For \pp\ this leads to $B\sim6\times10^{12}$\,G, which is, to some degree, high but still in the range of inferred dipole fields of rotationally-powered pulsars ($10^{11}-10^{13}$\,G). 
This pulsar is often compared to the Crab pulsar, another young and energetic object with a standard dipole field strength and similar rotational parameters, which also powers a luminous pulsar wind nebula (PWN).
However, the rotational behaviours of the Crab pulsar and \pp\ are nothing alike.

For about $16$\,yr, from February 1996 to late 2011, \pp\ had a stable rotational evolution with a braking index of $2.13\pm0.01$ \citep[]{2001ApJ...554L.177Z,2007Ap&SS.308..317L,2015ApJ...812...95F}. 
This regular spindown was only interrupted twice, by small glitches and their associated negative $\dot{\nu}$ steps (of $\sim0.01\%$ of $\dot\nu$) which had no visible recovery. Such behaviour is not unusual for pulsars with similar properties, and indeed, the Crab pulsar has mostly small to intermediate size glitches and otherwise evolves with a rather stable braking index close to $2.5$ \citep{2015MNRAS.446..857L}.  
Then in December 2011, extraordinarily, \pp\ began spinning down much faster than before, with $\dot{\nu}$ decreasing by about $36\%$ \citep{2015ApJ...807L..27M}. 
This enormous change was unexpected and did not appear related to a glitch: the spin frequency showed no change, and the measured $\Delta\dot\nu/\dot\nu$ was 10 to 100 times larger than the typical known $\dot\nu$ changes due to glitches.
Instead, the observed spin-down rate change could have been driven by magnetospheric changes, as there are reports of brightening of the PWN at the time of the spin-down event \citep{2019NatAs...3.1122G}. 
Unfortunately, there were no pulse profile or pulsed emission changes detected during this event (or any other time) which could confirm involvement of the magnetosphere \citep[][]{2015ApJ...812...95F,2015ApJ...807L..27M,2019NatAs...3.1122G,2024ApJ...967L..13T}.
Following the 2011 $\dot\nu$-drop, the braking index was considerably reduced: it was measured to be $n=0.03\pm0.01$ using a 480-days post-event dataset \citep{2016ApJ...827L..39M}, and $n=0.163\pm0.001$ using a 1100-days data span \citep{2019JKAS...52...41K}.
An increasing trend, from $n=0.1\pm0.1$ to $n=1.2\pm0.2$ was reported by \citet{2020MNRAS.494.1865W} analysing shorter shorter intervals over an extended 1500 days postevent data set.

More recently, \citet{2024ApJ...967L..13T} reported another rotational episode, which occurred in July 2023. 
The event was described as small decreases in both the frequency and the spin-down rate, i.e., an \emph{antiglitch} in spin frequency, but accompanied with a glitch-like change of the spin-down rate.
In this letter, we present the rotational evolution of \pp\ over the last $9.4$ years, from February 2015  until July 2024, describe its braking index behaviour, and re-analyse the 2023 event presenting a different interpretation.

\section{Instruments and Observations}

\label{obs_instr}

\pp\ was discovered in X-ray observations by \citet{1984ApJ...287L..19S} and has subsequently been detected in the optical, UV, gamma rays, and polarised X-rays \citep{1985Natur.313..659M,2019ApJ...871..246M,2015Sci...350..801F,2024ApJ...962...92X}. Its radio emission is very faint: $0.1$\,mJy in the MeerKAT L-band ($856$-$1712$\,MHz) \citep{1993ApJ...403L..29M,2003ApJ...590L..95J,2021MNRAS.505.4468G}, hence all timing observations of this pulsar are performed in the X-rays, most notably by the Rossi X-ray Timing Explorer \citep[RXTE,][]{2015ApJ...812...95F}. 
After the decommissioning of RXTE on January 2012, Swift-XRT \citep[][]{2005SSRv..120..165B} took on monitoring of \pp\ for about four years starting on February 2015 and ending on February 2019. 
This campaign started 3.1 years after the last RXTE observation, on Dec. 31, 2011; and is 
comprised of 78 observations performed at monthly cadence with typical exposure times of $0.5$-$2.5$\,ks, resulting in a total accumulated (screened) exposure time of $134.2$\,ks. 
The XRT operated in WT-mode providing a time resolution of 1.7675 ms, amply sufficient to study the timing behaviour of \pp.

NICER \citep[the Neutron Star Interior Composition Explorer Mission,][]{2016SPIE.9905E..1HG} started monitoring observations of \pp\ soon after its launch, on July 25, 2017, and it currently still executes observations of typically $0.5$-$2$\,ks at a weekly cadence.
The total accumulated screened exposure time up to and including July 29, 2024 is about $110.4$\,ks. 
The NICER monitoring observations show three large data gaps: I) Mar. 24, 2018 -- 
Mar. 20, 2019 , II) Jan. 22, 2021 -- Oct. 24, 2021 and III) Mar. 18, 2022 -- Jan. 19, 2023.
Fortunately the first gap had full coverage by concurrent Swift-XRT observations, so it is still possible to explore the timing characteristics during that period.
At the end of the third NICER data gap on Dec. 29, 2022 long polarimetric X-ray ($1$-$8$\,keV) observations by IXPE \citep[][]{2022JATIS...8b6002W} had been initiated. 
The last of the three observation clusters ended on May 12, 2023, overlapping with the NICER monitoring program. 
In total, $2.686$\,Ms of screened exposure time had been accumulated with IXPE.

\section{Timing analysis and results}

We used the aforementioned X-ray observations, from February 2015 to May 2024, to measure pulse arrival times (ToAs) according to the method outlined in Sect. 4.1 of \citet{2009A&A...501.1031K}. 
These ToAs were subsequently used for the timing analysis of the next two subsections.
Local fits to model the pulsar rotation close to a reference time $t_0$ are based on a Taylor expansion of the rotational phase, as 
\begin{equation}\label{phase1} 
\phi(t) = \nu_0 (t-t_0) + \frac{1}{2}\dot\nu_0 (t-t_0)^2 + \frac{1}{6}\ddot\nu_0 (t-t_0)^3 + \cdots 
\end{equation}
and were performed using the pulsar software {\sc TEMPO2} \citep{2006MNRAS.369..655H}.

\subsection{The braking index evolution}

\begin{figure}
\includegraphics[width=1.0\linewidth]{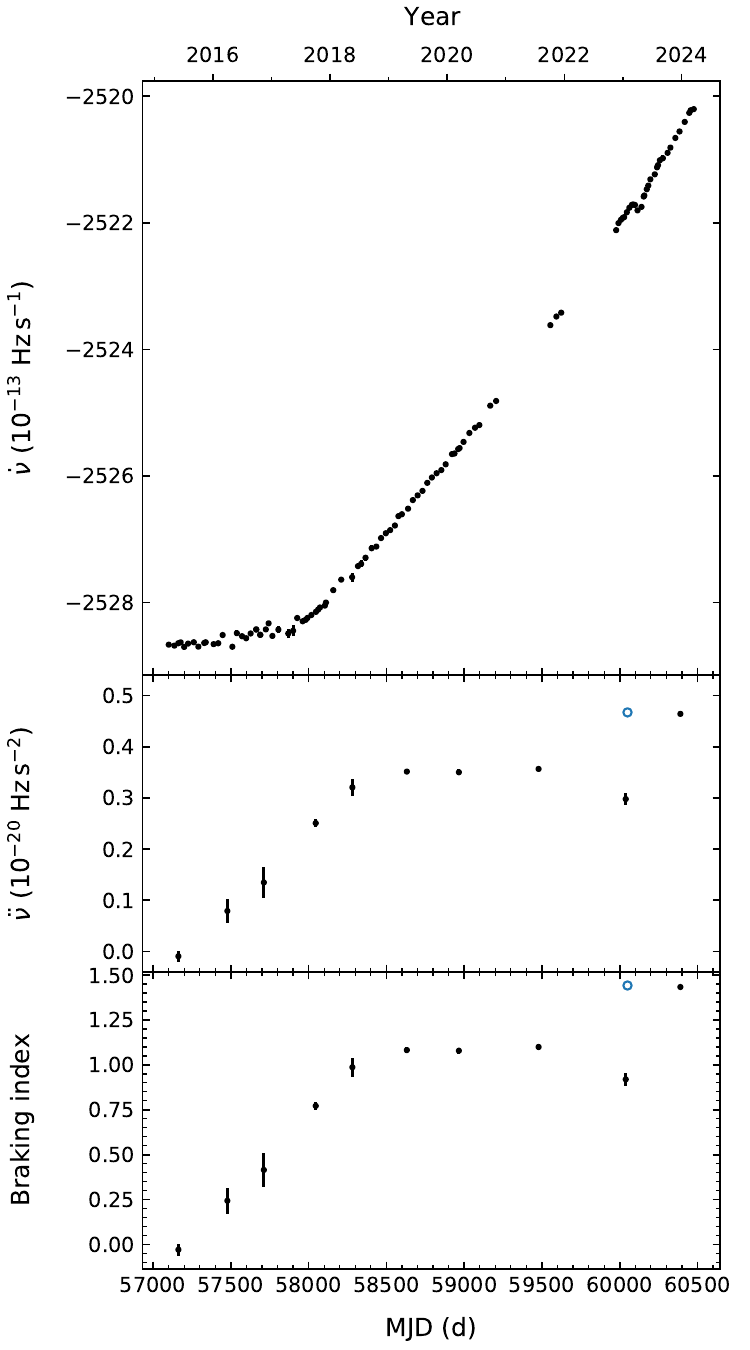}
\caption{The evolution of $\dot\nu$ of PSR B0540$-$69 between February 2015 and May 2024.
    \emph{Top panel}: The spin-down rate $\dot\nu$ measured from overlapping intervals approximately $80$\,d long, containing at least 6 TOAs each. 
    The data window was moved by $10$\,d at each stride and for these short fits $\ddot\nu$ was held constant, at a value measured from a larger TOA set (spanning at least $600$\,d).  
    \emph{Middle panel}: $\ddot\nu$ measured over $250$\,d long, non-overlapping, time intervals. For these fits, a third derivative of the frequency was included in Eq. \ref{phase1}, which was held as a fixed parameter at a value calculated from much longer intervals. These values are very small ($\dddot\nu<10^{-28}$\,Hz\,s$^{-3}$) and model a slow increase of $\ddot\nu$ that is observed across the studied dataset.
    \emph{Bottom panel}: The braking index $n=\nu\ddot{\nu}/\dot{\nu}^2$ calculated from each fit of the same intervals as in the middle panel.
    The blue, open circles in the bottom panels show measurements corrected by the 2023 timing event, according to the \emph{TN} model in Table \ref{tabla}.}
    \label{spinevo}
\end{figure}

Figure \ref{spinevo} illustrates the evolution of the spin-down rate $\dot{\nu}$, the second frequency derivative $\ddot{\nu}$, and the inferred braking index $n$ over the last 9.4 years. 
The datapoints are calculated from fits of Eq. \ref{phase1} over shorter time spans and, whilst their exact values are sensitive to the choice of parameters such as interval length and number of TOAs per fit (see figure caption for details), the results are qualitatively robust.

Initially, the spin-down rate evolves under a rapidly growing $\ddot{\nu}$ for about four years. 
During this phase, the braking index $n=\nu\ddot{\nu}/\dot{\nu}^2$ tracks the changing $\ddot{\nu}$ and increases from zero to $n\simeq1.1$ (middle and lower panels in Fig. \ref{spinevo}; see also \citet{2020MNRAS.494.1865W}). 
Unfortunately we cannot assess whether the rotation maintained an $n\sim0$ evolution between 2011 to 2015, or other developments had taken place. 
For example, a conjectural backwards extrapolation of the early trend seen in Fig. \ref{spinevo} would suggest a negative $\ddot{\nu}$ (and hence $n$) shortly before the onset of the observations presented here.

The period of increasing $\ddot{\nu}$ gave way to an approximately three years long period (2019, 2020, 2021), during which both $\ddot{\nu}$ and the braking index remain nearly constant. 
The change of $\ddot\nu$ during this stable phase is still positive, but very slow ($\dddot{\nu}\simeq(6\pm2)\times 10^{-30}$\,Hz\,s$^{-3}$). 
There are not enough observations available for 2022 to assess the stability of rotation during the second data gap (Fig. \ref{spinevo}), however, the first hundred days of the resumed observations by NICER at the start of 2023 suggest that the pulsar was evolving with a slightly larger braking index, of $1.4$ (not shown in Fig. \ref{spinevo}, which only displays results for $250$-d long intervals).

After that, in mid-2023, the pulsar underwent a rotational irregularity that cannot be described with a simple spin-down model as in Eq. \ref{phase1} and is the focus of the next subsection. 
After this latest event and towards the end of our dataset, the new rotational state exhibits $n=1.43\pm0.03$, but more data are necessary to establish the stability of this trend.

\subsection{The 2023 timing irregularity }

\begin{figure}
\includegraphics[width=1.0\linewidth]{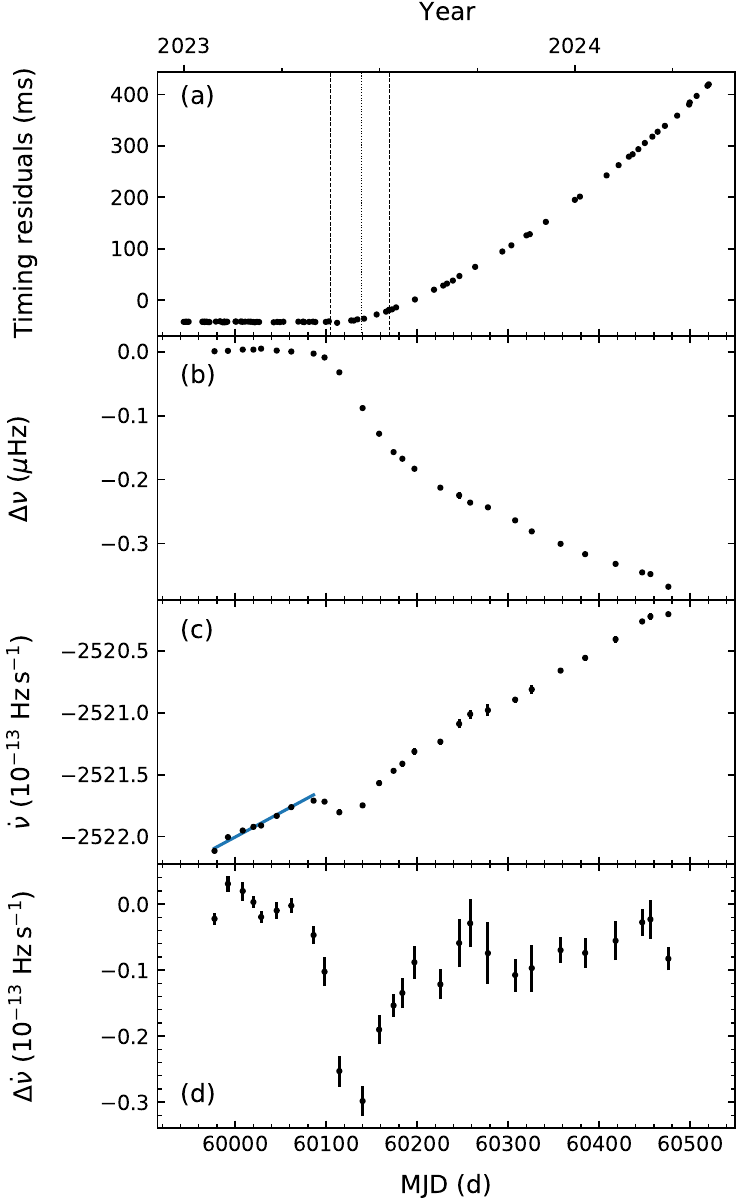}
\caption{The rotational evolution of PSR B0540$-$69 since January 2023.  
	\emph{(a)} The timing residuals relative to a model like Eq. \ref{phase1} adjusted up to MJD 60105.
		The two vertical segmented lines mark the epochs of the $\dot\nu$ changes according to the TN model ($t_{g1}$ and $t_{g2}$), while the vertical dotted line indicates $t_g$ according to the Glitch model.
	\emph{(b)} Frequency residuals relative to the same model.
	\emph{(c)} The behaviour of $\dot\nu$. The blue line represents the model used in the other panels.
	\emph{(d)} The residuals of $\dot\nu$ relative to the same model. 
	Frequency and spin-down rate values were calculated as in Fig. \ref{spinevo}, thus short term features ($<80$\,d) have been smoothed.
	\label{event}}
\end{figure}

The irregular behaviour shortly after MJD 60000 -- clearly visible in Fig. \ref{spinevo} -- was observed by NICER and first reported by \citet{2024ApJ...967L..13T}. 
The data and derived rotational parameters around this event are presented in Figure \ref{event}.
The top panel displays timing residuals (defined as the difference between the observed ToAs and the model prediction) with respect to the best-fit model of Eq. \ref{phase1} in the time interval between MJD $59936.3$ and MJD $60105.0$. 
The deviation from that timing model becomes apparent beyond that date, with the spin-frequency decreasing faster than predicted.

Including more terms (even up to ten time derivatives of the frequency) in Eq. \ref{phase1} still fails to describe the evolution after MJD 60105 and does not produce flat residuals dispersed around zero. 
Panels (b) and (d) of Fig. \ref{event} are illustrations of the frequency residuals and spin-down rate residuals with respect to the pre-event timing model, whilst panel (c) is the $\dot{\nu}$ curve. 
The data points for these three panels have been calculated as in Fig. \ref{spinevo}, from overlapping segments of TOAs, therefore fast changes on timescales shorter than $80$\,d have been smoothed out. 
Nonetheless, the plots aid to visualise the main features of the rotation before, during, and after the 2023 event.

\citet{2024ApJ...967L..13T} modelled the irregularity as if it were a glitch, with two discontinuities in $\nu$ and $\dot{\nu}$ at MJD 60132. 
Their result gives a negative frequency step $\Delta\nu<0$, and the event was named an anti-glitch because typical glitches have $\Delta\nu>0$ \citep{2022RPPh...85l6901A}. 
The spin-down rate change was found, however, to be negative ($\Delta\dot\nu<0$, as usual in ordinary glitches), which implies a progressive departure of the spin $\nu$ from its expected values. 
In the following, we re-evaluate the glitch model and present an alternative interpretation of the event, which is more strongly supported by the data. 

\begin{figure}
	\includegraphics[width=1\linewidth, trim=10 0 30 30, clip]{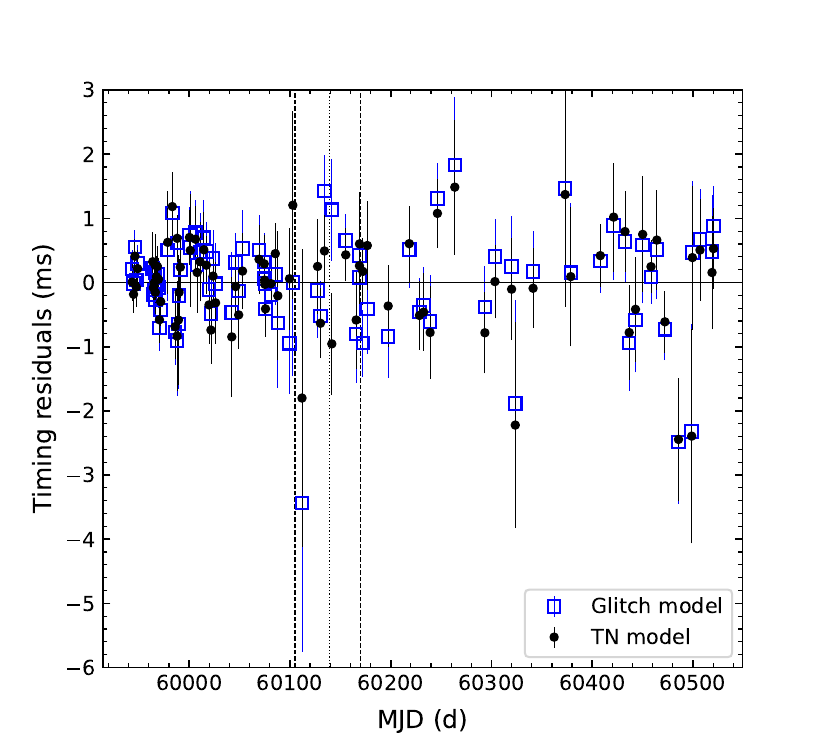}
	\caption{Timing residuals for the two models in Table \ref{tabla}.
	The vertical dotted line shows the epoch $t_g$ for the Glitch model, while the two vertical dashed lines mark the epochs $t_{g1}$ and $t_{g2}$ for the TN model.
		\label{resids}}
\end{figure}

A basic glitch model (that is, without including a post-glitch recovery) involves instantaneous, unresolved steps in spin frequency $\Delta\nu$ and spin-down rate $\Delta\dot\nu$, whose effects are added to the phase defined in Eq. \ref{phase1} after the glitch epoch, $t_g$:
\begin{equation}
\label{gmodel} 
\phi_g(t)=\Delta\phi + \Delta\nu (t-t_g) + \frac{1}{2}\Delta\dot\nu (t-t_g)^2 \quad, 
\end{equation}
in which $\Delta\phi$ ensures phase continuity across the event. 
Using our updated, longer dataset, we tested a glitch model as in Eq.~\ref{gmodel} but were unable to obtain flat timing residuals: obvious additional signal remains if only steps in $\nu$ and $\dot{\nu}$ are considered. 
To obtain an adequate fit, we must include an extra term describing a non-zero $\ddot{\nu}$ change in the model of Eq. \ref{gmodel}. 
The timing residuals of this model are shown as blue squares in Figure \ref{resids} and the best solution returns a frequency change of $\Delta\nu=-0.109\pm0.004$\,$\mu$Hz, in good agreement with \citet{2024ApJ...967L..13T}.

Further inspection of the data, however, reveals that this may not be the optimal description of the event.
A glitch leaves a cusp-like signature in the timing residuals --indicative of a discontinuity in spin frequency \citep{2011MNRAS.414.1679E,2022RPPh...85l6901A}, contrary to the timing residuals of \pp\ in this case, which show no indication of such feature (top panel in Fig. \ref{event}).
Instead, the frequency appears to change over a relatively short -- yet resolved -- timescale. 
This, combined with the fact that the spin-down rate $\dot\nu$ exhibits a relative decrease after MJD $\sim60100$ which almost completely reverts about 100 days later (panels (c) and (d) in Fig. \ref{event}), suggest that this event could be instead driven by a spin-down swing.

To test the above hypothesis, we develop and fit a second model, in which $\dot\nu$ changes discretely at two different times.
We call this the \emph{TN} model, as it resembles the timing noise of some pulsars in which abrupt $\dot\nu$ changes produce similar features \citep{2018MNRAS.475.5443S}. 
The \emph{TN} model does not involve any sudden frequency changes, hence $\Delta\nu=0$, and can be written as: 
%
\begin{multline}
\phi_{TN}(t)=\Delta\phi_1 + \frac{1}{2}\Delta\dot\nu_1 (t-t_{g1})^2 + \\
	H(t-t_{g2})\left(\Delta\phi_2+\frac{1}{2}\Delta\dot\nu_2 (t-t_{g2})^2 \right) \quad ,
\end{multline}
%
valid only for $t>t_{g1}$ and where $H$ is the Heaviside function so that the second spin-down step $\Delta\dot\nu_2 $ is added only after $t_{g2}$.
Discrete changes of the phase ($\Delta\phi_1$ and $\Delta\phi_2$, as for the glitch model,) associated with each event were also included, to ensure phase connection across each epoch. 
The residuals of this model are denoted with black circles in Fig. \ref{resids}.  

The following process was used to fit both the glitch and the \emph{TN} model to the data.
First, we used {\sc TEMPO2} to find the epochs ($t_g$ for the glitch model; $t_{g1}$ and $t_{g2}$  for the \emph{TN} model) that produced the smallest weighted Root Mean Square (RMS) of the timing residuals.
These epochs are quoted in Table \ref{tabla}.
In both cases the epoch(s) uncertainties are about $2$-$3$\,d, mostly reflecting the TOAs' typical separation in time. 
Whilst both models provide acceptable residuals, the RMS for the \emph{TN} model ($445$\,$\mu$s) is somewhat smaller than the one for the glitch model ($506$\,$\mu$s). 
This can also be detected by eye in Fig. \ref{resids}, however such a difference is not large enough to firmly favour one model over the other. 
We note though that an f-test gives a probability of $9\times 10^{-6}$, which could be an indication that the extra parameter of the \emph{TN} model is justified.
The $\chi^2$ and degrees of freedom (d.o.f.) for each fit are quoted in Table \ref{tabla}.

To draw conclusions on the preferred model, we then used a Bayesian approach to data fitting and compared the models' evidence.
For this, we used the timing software package PINT \citep{2021ApJ...911...45L} together with the {\tt nestle} library\footnote{https://github.com/kbarbary/nestle}, which uses Bayesian evidence and a nested sampling algorithm to explore the posterior distributions based on \citet[][]{2009MNRAS.398.1601F}. 
We used 5,000 points and an evidence precision of $10^{-5}$. 
The event epochs were held constant at the best values found before.
The most likely parameter values and their 68\% confidence intervals are reported in Table \ref{tabla}.
When comparing the final evidence ($Z$) of the glitch model over the \emph{TN} model, a Bayes factor of $3.8\times10^{-6}$ is obtained. 
This result implies a strong preference for the \emph{TN} model compared to the Glitch model.

\begin{deluxetable}{lll}
\tablecaption{Inferred timing parameters of the 2023 rotational irregularity for the two models considered. 
	Errors correspond to $68\%$ confidence intervals of the posterior distributions. 
	\label{tabla}}
\tablewidth{0pt}
\tablehead{	\colhead{Parameter} & \colhead{Glitch model} & \colhead{\emph{TN} model} 
}
\startdata
Epoch  (MJD)   							&	60050.0			& 60050.0		\\
$\nu$ (Hz)								&	19.635893779(1)	& 19.635893780(1)	\\
$\dot\nu$ ($10^{-10}$\,Hz\,s$^{-1}$)	&	-2.521821(2)	& -2.521803(2)	\\
$\ddot\nu$ ($10^{-21}$\,Hz\,s$^{-2}$)	&	4.2(1)			& 4.67(2)		\\
Event 1 (MJD)						    &	60139			& 60105		\\                  
$\Delta\phi_1$							&	0.01(1)			& 0.00(1)	\\
$\Delta\nu_1$ ($\mu$Hz)					&	-0.109(4)		& --		\\
$\Delta\dot\nu_1$ ($10^{-15}$\,Hz\,s$^{-1}$)	&	-5(1)	& -31.2(4)		\\
$\Delta\ddot\nu_1$ ($10^{-22}$\,Hz\,s$^{-2}$) 	&	 6(1)	& --		\\         
Event 2  (MJD)							        &	--		& 60170		\\         
$\Delta\phi_2$									&	--		& 0.02(1)		\\                
$\Delta\dot\nu_2$ ($10^{-15}$\,Hz\,s$^{-1}$)	&	--		& 21(1)		\\
RMS ($\mu$s)									&	505.85	& 445.45		\\
$\chi^2$/d.o.f.                                 &   94.8/79 & 73.5/78    \\
$\log(Z)$										&	591.24	& 603.71		\\
\hline
$Z_{\rm Glitch}/Z_{\emph{TN}}$   & \multicolumn2c{$3.8\times10^{-6}$} \\                               
MJD range                        & \multicolumn2c{59943.45 - 60520.27} \\                      
Number of ToAs                   & \multicolumn2c{88} \\                                 
\enddata
\tablecomments{Event 1 epoch refers to $t_g$ in the glitch model and to $t_{g1}$ in the \emph{TN} model. Event 2 refers to $t_{g2}$ in the \emph{TN} model.
	The uncertainties for these values are $2$-$3$\,d (see the text for more details).}
\end{deluxetable}

\section{Discussion}

While the \emph{TN} model may at first seem unusual, it has the potential of characterising timing irregularities in terms of physical processes that are often unexplored and different from glitches. 
Glitches have characteristic signatures and well established shared properties that are best explained by angular momentum transfer from an interior neutron superfluid to the rest of the star, causing it to spin-up \citep{1975Natur.256...25A}. 
On the other hand, the few reported antiglitches are typically in pulsars of high magnetic field and magnetars, each with its unique signature
\citep{2022RPPh...85l6901A}. These events are often associated with emission variations, indicating a possible magnetospheric origin or, at least, involvement. 
Smaller amplitude glitch-like and antiglitch-like anomalies, with $|\Delta\nu|<0.1$\,$\mu$Hz, are seen in the Crab and Vela pulsar (which have inferred magnetic fields of the order $10^{12}\;$G) and constitute a different population from their glitches \citep{2014MNRAS.440.2755E,2021A&A...647A..25E}.

The \emph{TN} model invokes rather fast changes in the torque without the need for an instantaneous frequency decrease. 
Such abrupt, transient, changes of $\dot{\nu}$, or even routine switches between different values, have been seen in several pulsars \citep[e.g., PSR B1822$-$09;][]{2010MNRAS.402.1027H,2018MNRAS.475.5443S}. 
They are mostly studied when they correlate with detectable changes in the radio pulse profile \citep{2010Sci...329..408L}, and can potentially explain part of the timing noise seen in all pulsars. 
Indeed, \citet{2021A&A...647A..25E} showed that several irregularities in the rotation of the Vela pulsar can be described with minute, sudden, changes just in $\dot{\nu}$ (with $\Delta\nu=0$). 
Moreover, \pp\ has already displayed a significant spin-down change in the past, hence it is not improbable that these smaller spin-down changes were caused by a similar physical process as the 2011 event.

We have also analysed one year of RXTE observations from 2003, and found a similar episode to the 2023 one, in which $\dot\nu$ at some point becomes smaller than the projection determined by $\ddot\nu$. 
The 2003 event is close to the second of the small glitches reported by \citet{2015ApJ...812...95F}, for which the measured changes $\Delta\dot\nu$ are similar in magnitude to those we find for the 2023 timing irregularity. 
Our preliminary analysis indicates that the 2003 event can also be well modelled by a \emph{TN} model. 
Is it possible that the two small glitches were instead $\dot\nu$ changes, rather than real glitches? 
We will address this question in future work, which will also aim to uncover the overall rotational history of \pp\ using archival data.

The relative magnitude of the 2011 $\dot{\nu}$ change is closer to those seen in intermittent pulsars \citep{2006Sci...312..549K,2013MNRAS.429.2569Y,2015ApJ...807L..27M}, 
except that so far $\dot{\nu}$ has not returned to its previous value, and might never do so if \pp\ reached a new steady state.
In a sense, the collection of peculiar $\dot{\nu}$ behaviours from \pp\ resembles more the erratic behaviour of magnetars.
In fact, the pulsar's current trajectory in the $P$-$\dot P$ diagram is towards the magnetars region, as determined by the current $n<2$. 
Note that this trend commenced dramatically with the large 2011 $\dot\nu$ change, when the inferred $B$ increased by about $10^{12}$\,G (and the characteristic age ($-\nu/2\dot\nu$) decreased by $440$\,yr). 
Unfortunately, it is not possible to use radio band observations of \pp\ to better understand its torque variability and its radio emission stability.

Processes typically put forward to explain low braking indices like those observed for \pp\ in its early stable period (until 2011) include significant contributions to spin-down by outflows, or a time-varying $k$ in Eq. \ref{powerlawspindown}. 
Such changes can arise, for example, due to magnetic field evolution in the neutron star crust or a decreasing effective moment of inertia as -- early in the star's life -- the interior neutrons turn superfluid. 
Sharp transitions though, as the large $\dot{\nu}$ drop in 2011, are most likely caused by magnetospheric variations, such as in the plasma outflow, or the overall structure e.g., the size of the open field line region. \citet{2015ApJ...807L..27M} explore some of these mechanisms for the \pp\ case. 
It is unfortunate that there is a time period of more than 1000 days without data after the 2011 giant $\dot\nu$ drop, which leaves unclear if and how the steady increase of $\ddot\nu$ and the evolving braking index shown in Fig. \ref{spinevo} are somehow connected to it.

Finally, note that the behaviour described for the braking index is determined by $\ddot\nu$ values measured over time spans which are considerably shorter than the timescale of the pulsar rotational evolution. 
The underlying longer-term trend is therefore still unclear. 
If anything, the enormous 2011 decrease in $\dot\nu$ has made the (discontinuous) long-term $\ddot\nu$ very small, if not negative.

\section{Final remarks} 

We have demonstrated that the 2023 rotational episode seen in \pp\ can be successfully modelled either as negative steps in both frequency and spin-down rate, plus an increase in $\ddot\nu$, or as a transient change of the spin-down rate.
The latter model, however, presents higher Bayesian evidence.  
Thus it may be common for \pp\ to experience variations of its spin-down rate, in the form of rapid changes in $\dot{\nu}$, or in the form of a highly varying $\ddot\nu$, or both. 
The presence of such rotational irregularities in \pp\ presents a great study opportunity because of its strong $\ddot{\nu}$ that can be accurately determined in relatively short timescales. 
Based on knowledge derived from radio pulsar observations, it is likely that the torque variations reflect magnetic reconfigurations or significant changes in the magnetospheric plasma density and outflow. 
Interestingly, \pp\ also shows an unexplained steady evolution of $\ddot{\nu}$ over a timescale of at least $\sim4$ years, which resulted in its braking index changing from nearly zero to a bit over $1$, though during and after the 2023 episode the new $n$ approaches $1.5$.
All these rotational features are not common in typical pulsars, thus it is of utmost importance to regularly monitor \pp\ by X-ray observatories, not only to accurately follow its rotation but also its emission in order to understand the role of the magnetosphere in the observed evolution and track any variations. To that end, continued observing in the radio (even if infrequent) is also imperative, especially shortly after the detection of any more rotational irregularities.


\begin{acknowledgments}
We are grateful for the computing facilities made available by ANID Chile SIA, grant SA77210112.
CME acknowledges support from ANID/FONDECYT, grant 1211964.
W.C.G.H. acknowledges support through grants 80NSSC23K0078 and 80NSSC24K1195 from NASA.
D.A. acknowledges support from an UKRI fellowship (EP/T017325/1). 
Astrophysics research at the Naval Research Laboratory is supported by the NASA Astrophysics Explorer Program.
\end{acknowledgments}

%

\vspace{5mm}
\facilities{Swift(XRT), NICER, IXPE}


\software{{\sc TEMPO2} \citep{2006MNRAS.369..655H},
	{\sc PINT} \citep{2021ApJ...911...45L},  
    nestle (\url{https://github.com/kbarbary/nestle})          
}


\bibliography{export-bibtex}{}
\bibliographystyle{aasjournal}



\end{document}